\documentclass[11pt,twoside]{article}
\usepackage{asp2004}
\usepackage{psfig}
\usepackage{epsf}
\usepackage{graphics}
\usepackage{lscape}
\begin{document}
\title{The Spreading Layer and Dwarf Nova Oscillations}
\author{Anthony L. Piro}
\affil{Department of Physics, Broida Hall, University of California,
Santa Barbara, CA 93106}
\author{Lars Bildsten}
\affil{Kavli Institute for Theoretical Physics and Department of Physics,
Kohn Hall, University of California, Santa Barbara, CA 93106}

\begin{abstract}
We describe recent theoretical work on the final stage of
accretion when material passes from an accretion disk onto
a white dwarf surface. Our calculations focus on understanding
the latitudinal spreading and differentially rotating profile
of this material, so we call it the ``spreading layer'' (SL)
model. The SL typically extends to an angle of
$\theta_{\rm SL}\approx0.01-0.1$ radians, with respect to the equator.
At low accretion rates ($\dot{M}\la10^{18}\textrm{ g s}^{-1}$)
the amount of spreading is small, so that the dissipated
energy is radiated back into the accretion disk. When $\dot{M}$
is high, such as in dwarf novae, symbiotic binaries, and
supersoft sources, the material
spreads to higher latitudes to be more easily observed. The SL may
contain shallow surface modes, and we propose that such waves could
produce dwarf nova oscillations (DNOs). This hypothesis
naturally explains many key properties of DNOs, including
their frequency range, sinusoidal nature, sensitivity to $\dot{M}$,
and why they are only seen during outburst.
\end{abstract}
\thispagestyle{plain}

\section{Introduction}

  If a white dwarf (WD) is weakly magnetized, accreted material
reaches its surface at the equator and then must lose angular
momentum and kinetic energy before settling onto the star. When
the WD spin is much less than breakup, nearly half of the accretion
luminosity is released in this process, making this region as
luminous as the accretion disk and crucial to understanding the
luminosity from cataclysmic variables (CVs). This transition from accretion
disk to WD is typically treated in a boundary layer (BL) model.
One-dimensional BL studies follow the radial flow of material and
assume a vertical scale height (e.g. Popham \& Narayan 1995), while
other BL models study both radial and latitudinal directions
simultaneously using numerical techniques 
(e.g. Kley 1989a, 1989b). Inogamov \& Sunyaev (1999; hereafter IS99)
approached this problem from a new angle to study accreting neutron stars
(NSs). Their method follows the latitudinal flow of matter and
provides information about the spreading area of hot, freshly
accreted material which is not captured in BL models. We apply these
same methods to the case of WDs (Piro \& Bildsten 2004a; hereafter PB04a)
and call this model the spreading layer (SL), to differentiate it from
BL studies.

  The SL is found to be a thin hot band of extent
$\theta_{\rm SL}\approx0.01-0.1$ radians, with an effective temperature
of $\sim(2-5)\times10^5\textrm{ K}$ for
$\dot{M}=10^{17}-10^{19}\textrm{ g s}^{-1}$,
implying that it contributes to accreting WD spectra in the extreme
ultraviolet (EUV) and soft X-rays. At low $\dot{M}$
($\la10^{18}\textrm{ g s}^{-1}$) the spreading is small, so that its
radiation is absorbed by the accretion disk or its winds. Nevertheless,
the SL may be important for calculating the underlying continuum
scattered by these winds.

    The material in the SL is much hotter in temperature and lower in
density than the underlying WD. This contrast allows waves
in the SL to travel freely, unencumbered by the material below. We
propose that dwarf nova oscillations (DNOs) are shallow surface waves
in a layer of recently accreted material confined to the WD equator
(Piro \& Bildsten 2004b; hereafter PB04b). This model explains the
main properties of DNOs, including their sinusoidal nature,
dependence on $\dot{M}$, large pulsed amplitude in the EUV,
and why they are only seen during outburst.

\section{The Spreading Layer Calculation}

  The SL is treated as a one-zone, plane-parallel layer
with $\theta$ (the angle from the equator) as the independent
coordinate. The surface gravitational acceleration is significantly
decreased by centrifugal effects due to the nearly Keplerian speed
of the layer, so we use an ``effective'' gravitational constant
\begin{eqnarray}
        g_{\rm eff} = \frac{GM}{R^2}
                - \frac{v_\theta^2}{R}
		- \frac{v_\phi^2}{R}
		\approx \frac{GM}{R^2}
                - \frac{v_\phi^2}{R}
\label{eq:geff}
\end{eqnarray}
where $v_\theta$ and $v_\phi$ are the latitudinal and azimuthal velocities,
respectively. Hydrostatic balance integrates to $P=g_{\rm eff}y$, where
$y$ is the column depth (cgs units of $\textrm{g cm}^{-2}$).
The flux in the layer is assumed constant
since most of the viscous dissipation occurs at the layer's
base. Radiative diffusion integrates to
\begin{eqnarray}
	F = \frac{acT^4}{3\kappa y},
\label{eq:flux}
\end{eqnarray}
where $a$ is the radiation constant and
$\kappa=0.34 \textrm{ cm}^{2}\textrm{ g}^{-1}$ is
the Thomson scattering opacity for a solar composition.

  We insert this one-zone model into a set of differential equations
describing the steady-state conservation of mass, momentum,
and energy, as done by IS99 for NSs.
The total viscous stress, $\tau$, is from the turbulence created as high
entropy fluid quickly rotates against the lower entropy WD material below,
so parametrizing $\tau$ in terms of a unitless constant, $\alpha_{\rm SL}$,
\begin{eqnarray}
        \tau = \alpha_{\rm SL} \rho v^2,
\end{eqnarray}
where $v^2 = v_\theta^2+v_\phi^2$. We treat
$\alpha_{\rm SL}$ as a free parameter and study its effect
on the spreading properties
for the values $\alpha_{\rm SL}=10^{-4}-10^{-2}$. This range is chosen for
the physically realistic solutions that result (see PB04a for further
discussion).

  Integration requires setting boundary conditions for $\theta$, $T$,
$v_\theta$, and $v_\phi$. Shakura \& Sunyaev (1973; for gas pressure
dominating radiation pressure and a Kramer's opacity) show the disk
subtends an angle at the WD surface
\begin{eqnarray}
	\theta_{\rm disk} = 1.8\times10^{-2}
        \alpha_{\rm disk,2}^{-1/10}
        \dot{M}_{17}^{3/20}
        M_1^{-3/8}
        R_9^{1/8}
\label{eq:thetadisk}
\end{eqnarray}
where $\alpha_{\rm disk}$ is the viscosity parameter for the accretion disk,
$\alpha_{\rm disk,2}\equiv\alpha_{\rm disk}/10^{-2}$,
$\dot{M}_{17}\equiv\dot{M}/(10^{17} \textrm{ g s}^{-1})$,
$M_1\equiv M/M_\odot$, $R_9\equiv R/(10^9 \textrm{ cm})$,
and we set the factor $[1-(r/R)^{-1/2}]\approx1$. The
material which comes in from the disk must start at an angle
$\la\theta_{\rm disk}$,
so we use an initial angle of $\approx10^{-3}$. The initial
$v_\phi$ is set to $0.99v_{\rm K}$, where $v_{\rm K}=(GM/R)^{1/2}$ is the
Keplerian velocity, and the initial $T$ is set using the disk's midplane
temperature (Shakura \& Sunyaev 1973).
Finally, the initial $v_\theta$ is set by
the natural tendency for the differential equations to asymptote to
the limit where initially the viscous dissipation
is radiated away locally (PB04a). This is different than
what IS99 find for NSs, where more advection takes place, due to
$\dot{M}$ closer to the Eddington limit.

\begin{figure}[!ht]
\plotone{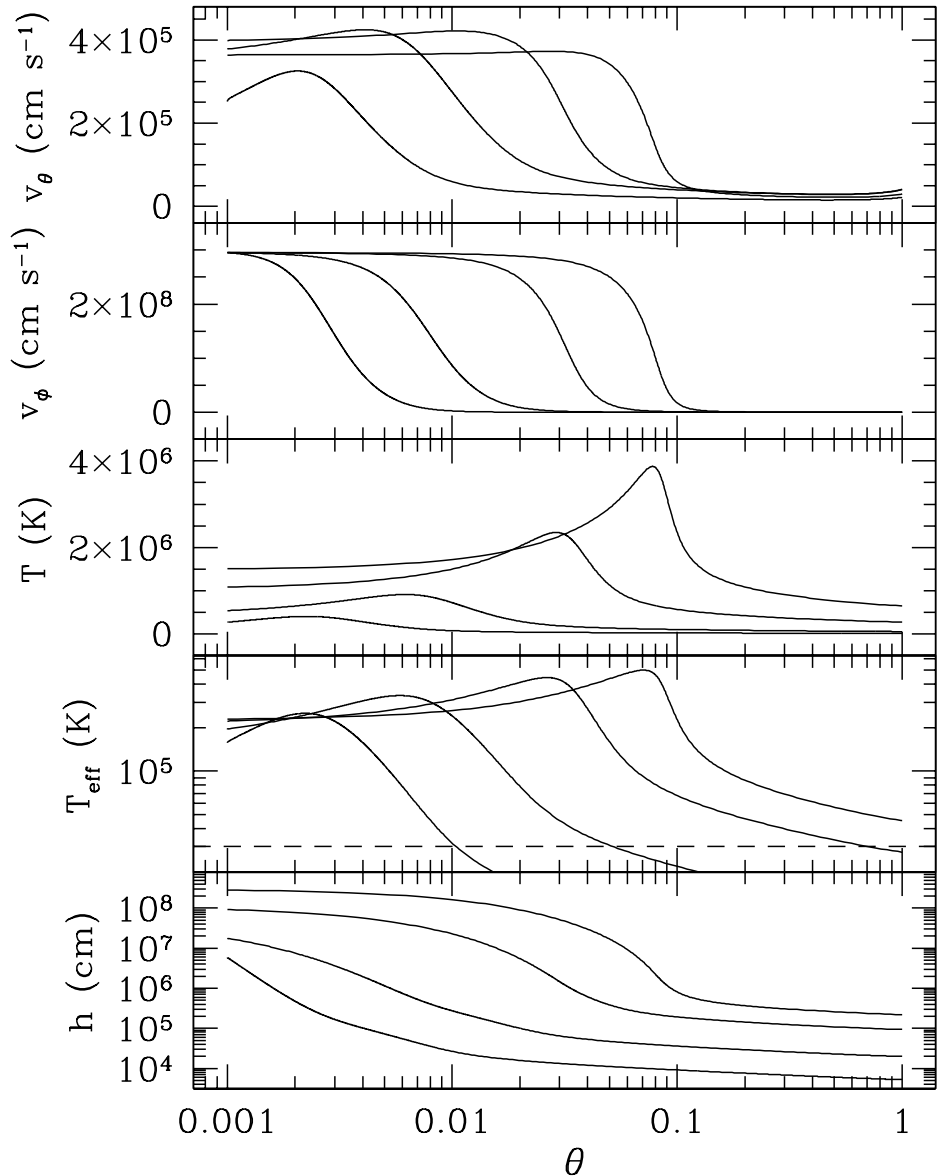}
\caption{The velocities, temperature, effective temperature and
pressure scale height of the SL
for a WD with $M=0.6M_\odot$ and $R=9\times10^8\textrm{ cm}$,
with $\alpha_{\rm SL}=10^{-3}$ and $\alpha_{\rm disk}=0.1$.
From left to right the $\dot{M}$ are $10^{17}$, $10^{18}$, $10^{19}$ and
$3\times10^{19}\textrm{ g s}^{-1}$. The dashed lines shows
$T_{\rm eff}=3\times10^4\textrm{ K}$, a fiducial temperature for the
underlying accreting WD (Townsley \& Bildsten 2003).}
\label{fig:slprofile1}
\end{figure}
 In Figure \ref{fig:slprofile1} we plot the velocities, $T$, $T_{\rm eff}$ and
the pressure scale height, $h=P/(\rho g_{\rm eff})$, of the SL for
$\alpha_{\rm SL}=10^{-3}$ and a range of $\dot{M}$. At the high accretion rates
($\dot{M}\ga10^{19}\textrm{ g s}^{-1}$) more advection takes place,
resulting in solutions which look similar to what IS99 found for
NSs. Most notably, the $T$ profiles show large peaks
at high latitudes due to energy being advected away from the equator and
deposited higher up on the star. The panel plotting $T_{\rm eff}$ has
an additional dashed line at $3\times10^{4}\textrm{ K}$, a fiducial
temperature of the underlying accreting WD
(Townsley \& Bildsten 2003) to show the contrast between it and the SL.

  Using the results from Figure 1, we plot the dependence of
$\theta_{\rm SL}$ on $\dot{M}$ in Figure 2. This shows
that high $\dot{M}$ are needed for any hope of appreciable spreading
to take place, so that typically the SL will be absorbed and re-radiated by
the accretion disk and its winds. Nevertheless, even when the accretion
disk covers the main portion of the SL, the profiles of $T_{\rm eff}$ shown
in Figure 1 are important for understanding the WD spectra.
\begin{figure}[!ht]
\plotone{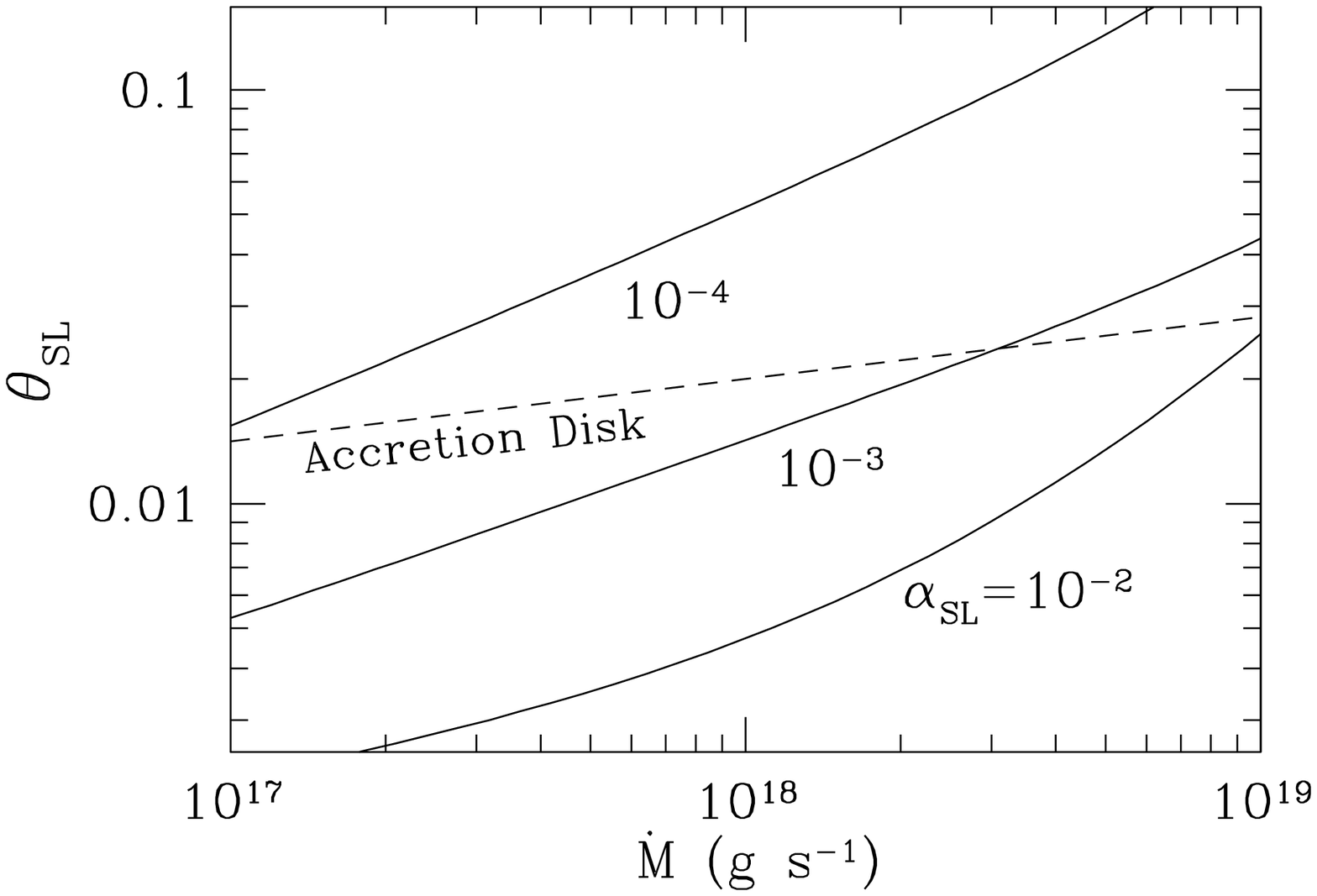}
\caption{The spreading angle, $\theta_{\rm SL}$, versus $\dot{M}$
for a WD with $M=0.6M_\odot$ and $R=9\times10^9\textrm{ cm}$
and three values of $\alpha_{\rm SL}$. The spreading angle is
defined as the angle at which $v_\phi=0.1v_{\rm K}$.
This shows that high $\dot{M}$ are needed for the
SL to extend beyond the angle made by a Shakura-Sunyaev disk,
equation (4) (dashed line).}
\end{figure}

\section{Comparisons to Dwarf Novae in Outburst}
  
  The high $\dot{M}$ needed for appreciable spreading
suggests that a good candidate to show a SL are dwarf novae
in outburst.
Evidence for spreading can be explored by following $T_{\rm eff}$
versus $\dot{M}$ in the EUV. Since we find that the
radiating area increases with $\dot{M}$, the SL model predicts a
shallower change in $T_{\rm eff}$ with $\dot{M}$ than if the radiating
area was fixed. PB04a use this relation in an attempt test for
evidence of spreading during an outburst of SS Cyg
(Wheatley, Mauche \& Mattei 2003), but are not able to make any
definite conclusions with the current data. Further detailed observations
following CVs in outburst
are needed to test the SL model in better detail.

  The ease of investigating the SL depends on the angle of the observed
system. Face-on systems show a continuum most
likely from the SL region, but with strong absorption lines from the
disk winds (Mauche 2004). Assuming the fractional emitting area is
approximately equal to $\theta_{\rm SL}$,
modeling of this spectra finds
$\theta_{\rm SL}\approx5.6^{+60}_{-4.5}\times10^{-3}$ (for SS Cyg;
Mauche 2004),
which, though not highly constraining,
is consistent with our calculations.
Edge-on systems show evidence of a SL emission that is
scattered into the line of sight by line emission in the disk winds
(OY Car; Mauche \& Raymond 2000), but is
otherwise obscured by the accretion disk.
It is therefore
difficult to make direct comparisons with the SL model.

\section{Dwarf Nova Oscillations}

DNOs are oscillations observed during dwarf nova outbursts that 
have periods of $P\approx3-40\textrm{ s}$, scale monotonically
with EUV luminosity, and thus $\dot{M}$ (Mauche 1996), and
are fairly coherent ($Q\sim10^4-10^6$). Their large pulsed fraction
in the EUV implies that they originate close to the WD surface and
have a similar covering fraction as the SL (C\'{o}rdova et al. 1980;
Mauche \& Robinson 2001; Mauche 2004).
We propose that DNOs are produced by nonradial oscillations
in the SL (PB04b). In the following sections we present a simple model to
describe these oscillations and make comparisons with observations.

\subsection{Nonradial Oscillations in the SL}

  Patterson (1981) showed that (at the time) all known DNOs on WDs with
measured masses have $P\ga P_{\rm K}=2\pi(R^3/GM)^{1/2}$.
This relation is an important constraint for any explanation of DNOs,
so we first show why a SL mode should mimic such a period.
The large entropy contrast between the surface layer and the underlying
material confines nonradial oscillations to high altitudes with little
or no pulsational energy extending
deeper into the WD. When the horizontal wavelength is much greater than
the layer depth, these shallow surface waves have a frequency
\begin{eqnarray}
        \omega^2 = g_{\rm eff}hk^2,
\end{eqnarray}
where $k$ is the transverse wavenumber. The SL can be thought of as a
waveguide with latitudinal width $2\theta_{\rm SL}R$ and azimuthal
length $2\pi R$, but since
$\theta_{\rm SL}\ll1$ the latitudinal contribution dominates, so
$k\sim1/(\theta_{\rm SL}R)$. Setting $g_{\rm eff}=\lambda GM/R^2$, where
$\lambda\la1$ is a dimensionless parameter that depends
on the spin of the layer, we rearrange the terms in equation (5) to find
\begin{eqnarray}
        \omega =
                \left( \frac{GM}{R^3} \right)^{1/2}
                \left( \frac{\lambda h}{\theta_{\rm SL}^2R} \right)^{1/2},
\end{eqnarray} 
so that the mode's frequency is the Keplerian frequency times
a factor less than unity (as long as $\lambda h\la \theta_{\rm SL}^2R$), and
therefore consistent with Patterson (1981).

  To find how these modes scale with $\dot{M}$ and $M$ we
consider a simple model to estimate $h$. Approximately half of
the accretion luminosity is released at the WD surface, so that
the flux of this layer is given by
$4\pi\theta_{\rm SL}R^2 F = GM\dot{M}/(2R)$. This
flux is related to $T$ by radiative diffusion, equation (\ref{eq:flux}).
The column depth is set by continuity to be
$y=\dot{M}t_{\rm SL}/(4\pi \theta_{\rm SL}R^2)$,
where $t_{\rm SL}=h^2/\nu$ is the timescale for viscous dissipation
in the SL. The viscosity between the fresh material and the
underlying WD is given by
$\nu\approx\tau h/(\rho v_\phi)\approx\alpha_{\rm SL}v_\phi h\approx\alpha_{\rm SL}v_{\rm K} h$.
Using an ideal gas equation of state, the above set of equations
that can be solved for $h$.
We set $\theta_{\rm SL}\sim\theta_{\rm disk}$ using
equation (\ref{eq:thetadisk}), since the disk may be setting the
spreading angle as suggested by Figure 2.
Equation (5) then gives a period
\begin{eqnarray}
        P_{m=0} = 30\textrm{ s }\alpha_{\rm disk,2}^{-2/15}
                \alpha_{\rm SL,3}^{1/6}\lambda_1^{1/6}\dot{M}_{17}^{-2/15}
                M_1^{-1/3}R_9^{19/12},
\end{eqnarray}
where $\lambda_1\equiv\lambda/10^{-1}$. Since this lowest order mode
does not propagate in the azimuthal direction, we denote it with azimuthal
wavenumber $m=0$.
Equation (7) provides both a scaling with $\dot{M}$
and a period suggestive of DNOs.

  The SL can contain additional modes,
most notably those that propagate in
the azimuthal direction. These modes have an
observed frequency of $\omega_{\rm obs}=|\omega-m\omega_{\rm SL}|$, where
$\omega_{\rm SL}$ is the spin of the SL and $m$ is the azimuthal
wavenumber. Even though the layer is spinning quickly,
Coriolis effects will not alter $k$, since
$\theta_{\rm SL}\la\omega/(2\omega_{\rm SL})$
(Bildsten, Ushomirsky \& Cutler 1996). Setting
$P_{\rm SL}=2\pi/\omega_{\rm SL}$, where $P_{\rm SL}\ga P_{\rm K}$,
the next lowest order modes have
\begin{eqnarray}
	\frac{1}{P_{m=\pm1}}
	= \left|\frac{1}{P_{m=0}}\mp\frac{1}{P_{\rm SL}}\right|.
\end{eqnarray}
for their periods.

\subsection{DNOs in the CV Population}
\begin{figure}[!ht]
\plotone{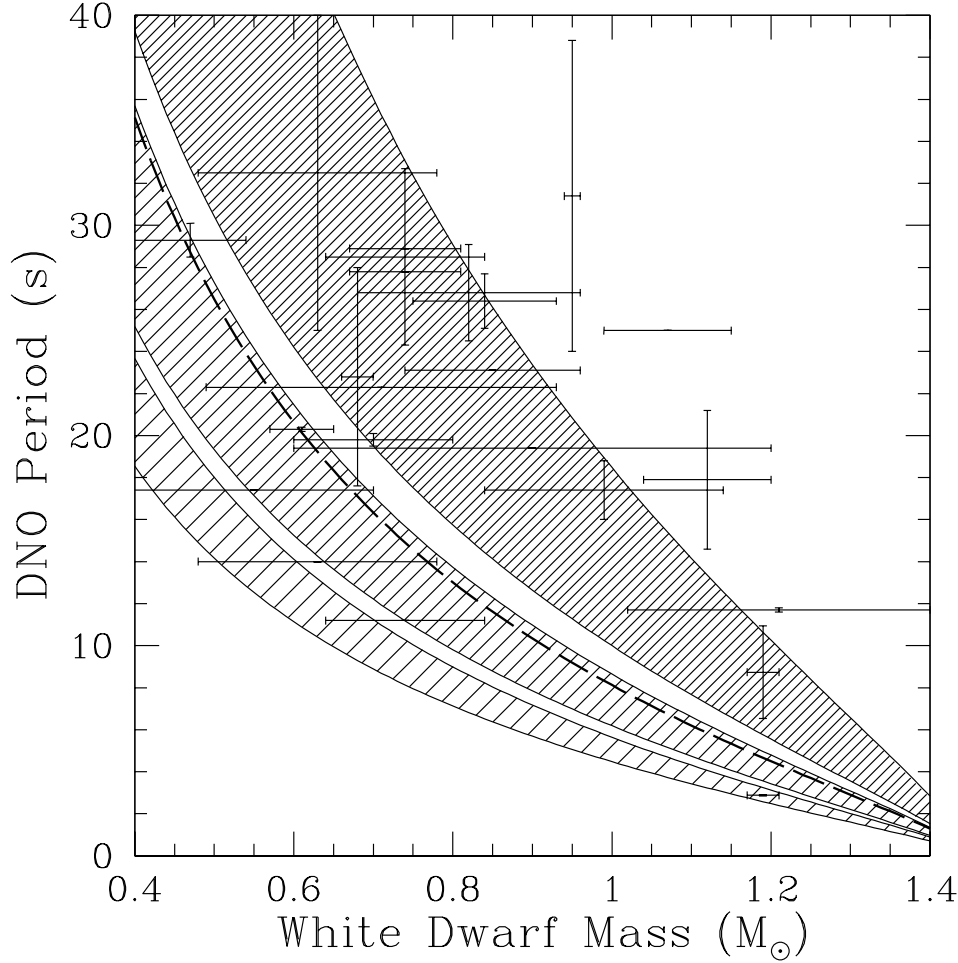}
\caption{The DNO periods versus measured WD masses. The vertical bars show
the range of periods observed from each CV, while the horizontal bars
are mass measurement errors. CVs that have shows two distinct DNO periods are
plotted as two separate points.
The thick dashed line is the Keplerian period, $P_{\rm K}$. 
The $m=0$ mode (wide, heavy shaded region) is plotted for
$\dot{M}=10^{16}-10^{18}\textrm{ g s}^{-1}$
(from top to bottom), using equation (7). The $m=-1$ mode
is also plotted for $\dot{M}=10^{16}-10^{18}\textrm{ g s}^{-1}$
(from top to bottom),
for both $P_{\rm SL}=P_{\rm K}$ (light shaded) and $P_{\rm SL}=2P_{\rm K}$
(medium shaded), using equation (8).}
\end{figure}
  We compare the most recent compilations of DNO periods (Table 1 of
Warner 2004) and WD masses (Ritter \& Kolb 2003) of CVs in Figure 3.
We plot systems that have shown both high and low DNO periods as
two separate points (these CVs are SS Cyg, CN Ori, and VW Hyi).
The heavy, dashed line denotes the surface Keplerian period, $P_{\rm K}$,
for a given WD mass (using the mass-radius relation of Truran
\& Livio 1986). This demonstrates that DNOs exist both above and
below $P_{\rm K}$, so
we consider SL nonradial oscillations with $m=0$ and $m=-1$.
For each mode we plot periods for
$\dot{M}=10^{16}-10^{18}\textrm{ g s}^{-1}$, the range
expected during a DN outburst. To calculate the
$m=-1$ mode we must assume a period for the SL's spin,
$P_{\rm SL}$, so we plot mode periods for both $P_{\rm SL}=P_{\rm K}$
and $P_{\rm SL} = 2P_{\rm K}$. This shows that our model is
insensitive to the SL spin rate.

  Besides predicting multiple classes of DNOs with different period
ranges, we also predict that each of these groups should have
a different dependence on $\dot{M}$. This can be seen from the size
of each shaded band, which is much wider for the $m=0$ mode than the
$m=-1$ mode. The DNOs with $P\la P_{\rm K}$ show less variation with
$\dot{M}$, qualitatively consistent with the $m=-1$ mode of our model.
To further investigate the $\dot{M}$ dependence we
plot our predicted DNO periods versus $\dot{M}$ for a $M=1.0M_\odot$
WD in Figure 4. This illustrates the shallower dependence for the
$m=-1$ mode. The separation and relative slopes of the $m=-1$ and
$m=0$ modes are suggestive of the frequency doubling
during the October 1996 outburst of SS Cyg (Mauche \& Robinson 2001).
\begin{figure}[!ht]
\plotone{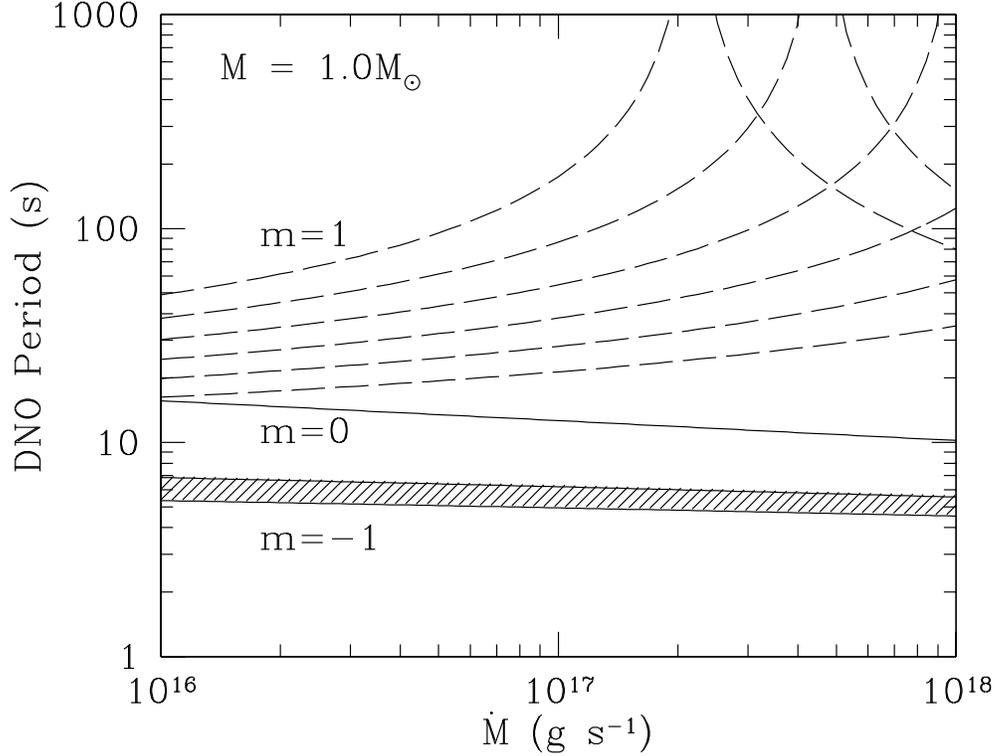}
\caption{The SL nonradial oscillation period compared to
$\dot{M}$ for a $M=1.0M_\odot$ WD.
The solid, heavy line is the $m=0$ mode, while the
shaded region shows the $m=-1$ mode for a range of
$P_{\rm SL}=(1.0-1.5)\times P_{\rm K}$ (bottom to top). The dashed lines
show the $m=1$ mode for the same range of $P_{\rm SL}$.}
\end{figure}

  In Figure 4 we also plot the $m=1$ retrograde mode for
$P_{\rm SL}=(1-1.5)\times P_{\rm K}$.
Due to the difference taken in equation (8), this mode has a
complicated dependence on $\dot{M}$, no longer being
a power law nor monotonic. It has a higher period than typical DNOs,
and may be relevant for the long period DNOS
(lpDNOs; Warner, Woudt \& Pretorius 2003).
To identify these as modes requires
checking for this predicted scalings between $P$ and $\dot{M}$.
Such a test may negate a mode explanation for lpDNOs since
Warner et al. (2003) claim to find no such
correlation and thus favor a spin related mechanism.
On the other hand, on separate occasions there have been
$32-36\textrm{ s}$
(Robinson \& Nather 1979) and $83-110\textrm{ s}$ (Mauche 2002b)
oscillations seen from SS Cyg, in the domain expected for these modes.
This wide
spread of periods may be explained by
the $m=1$ mode's steeper dependence on $\dot{M}$.
The $32-36\textrm{ s}$ oscillations showed less coherence than
typical DNOs, which could be due to these retrograde oscillations
beating against the accretion disk.
Oscillations from VW Hyi at $\sim90\textrm{ s}$
(Haefner, Schoembs \& Vogt 1977) may be of similar origin.

  An important property of WDs that would affect the modes is
a magnetic field. A strong field inhibits shearing between the
SL and WD and modifies the frequency of shallow surface
waves. It is therefore interesting that no intermediate polars
(IPs) have shown DNOs or lpDNOs. Even LS Peg and
V795 Her, neither of which are IPs, but both show polarization modulations
(Rodr\'{i}quez-Gil et al. 2001, 2002) indicative of a reasonably strong
magnetic field,
are without DNOs or lpDNOs.

\section{Discussion and Conclusions}

The SL model provides a new area of investigation in the study of
accreting objects. The initial calculations by IS99 and in this paper
describe the main features of spreading that will lead to
further studies of how accreted material settles onto stars. This may include
studying how differential rotation affects the underlying stellar surface.
Spectral modeling
of the SL, along comparisons with observations, will help in identifying
if and when spreading is present. In this review we focus
on comparisons with dwarf novae, but symbiotic binaries (Sokoloski 2004)
and supersoft sources are also promising for studying the SL.

  We propose that DNOs in outbursting CVs are nonradial oscillations
in the SL. A number of DNO properties are then simply understood:
(1) the highly sinusoidal nature of the oscillations is consistent with
nonradial oscillations, (2) the periods can change on the timescale of
accretion because
there is little mass in the layer ($\la10^{21}\textrm{ g}$; PB04),
(3) the periods vary inversely with $\dot{M}$ because they have the
temperature scaling of shallow surface waves, (4) the covering fraction is
naturally small for the SL, (5) the DNOs are only seen during DN outbursts
when an optically thick layer of material
builds up at the equator, and (6) the largest pulsed amplitude is in the
EUV, consistent with the SL temperature.

  There are difficulties that must be answered about this
idea for explaining DNOs. From the SL model we borrowed the
concept of hot material in hydrostatic balance covering a small fractional
area of the WD, but we ignored important details of these calculations,
and instead only presented a very simple model for the modes.
The data do not require such additional complications,
so we refrain from including them for now. In a more
sophisticated model it would probably still be true that
$k\propto1/(\theta_{\rm SL}R)$,
but an eigenvalue calculation would determine the
constant of proportionality.

  Further studies should work toward understanding
the excitation mechanism for the modes, which would explain
the high coherence typical of DNOs. Material deposited at the WD equator
spreads quickly, $\sim10-100\textrm{ s}$ (PB04a), so that following an
outburst, when $\dot{M}$ has fallen, the accreted material will spread over
the star and not be seen as a separate hot component.
This also means that the material is
moving through the oscillating region on timescales
of order the mode period. This short timescale is problematic if the modes
are to remain coherent, which puts limits
any proposed excitation mechanism.

  Our explanation of DNOs raises interesting questions about the
relationship between oscillations originating from accreting compact
objects. Mauche (2002b) showed that there is a correlation in
the high to low oscillation frequency ratio of WDs, NSs,
and black holes (BH). Using this picture, DNOs are associated with
the kilohertz QPOs of low mass X-ray binaries. Suggestively, the
Fourier frequency resolved spectroscopy of NSs
(Gilfanov, Revnivtsev \& Molkov 2003)
imply that both the normal branch oscillations and the kilohertz
QPOs originate in the NS BL. Using a NS mass and
radius our model provides frequencies in the range
expected for kHz QPOs, but it does not explain their
$P$-$L$ relation (the ``parallel
tracks''; van der Klis 2000). Interestingly, DNOs may
also show the parallel tracks phenomenon, as seen in three observations
of SS Cyg (Mauche 2002a), once again supporting the correlation.
On the other hand, continuing the analogy to
WD and BH oscillations is problematic because in the case of
BHs there is no surface for nonradial oscillations.

\acknowledgements{We thank Phil Arras, Philip Chang, Christopher Deloye,
Christopher Mauche, Aristotle Socrates, Rashid Sunyaev, and Dean Townsley
for many helpful conversations.
This work was supported by the National Science Foundation
under grants PHY99-07949 and AST02-05956, and by the Joint Institute
for Nuclear Astrophysics through NSF grant PHY02-16783.}

\end{document}